\documentclass[final]{aipproc}
\layoutstyle{6x9}
\usepackage{graphicx}
\usepackage{epsf}
\usepackage{wrapfig}
\usepackage{sublabel}
\usepackage{epsfig}
\usepackage{amsmath}

\newcommand{\beq}{\begin{eqnarray}}

\newcommand{\be}{\begin{equation}}
\newcommand{\ee}{\end{equation}}

\newcommand{\eeq}{\end{eqnarray}}

\def\b0{{\mbox{\boldmath$0$}}}

\def\b0{{\mbox{\boldmath$0$}}}

\def \b #1{ {\bf #1}}

\def \b #1{ {\bf #1}}

\def\Bf#1{\mbox{\boldmath $#1$}}

\def\beqy{\begin{eqnarray}}
\def\eeqy{\end{eqnarray}}

\def\mh{\hat{\mathcal{O}}}
\begin{document}
\vskip 2mm \date{\today}\vskip 2mm
\title{Short Range  Correlations in Medium- and
High-Energy Scattering off Nuclei}
\author{M. Alvioli}{
  address={Department of Physics, University of Perugia and
    Istituto Nazionale di Fisica Nucleare, Sezione di Perugia,
    Via A. Pascoli, I-06123, Italy}}
\author{C. Ciofi degli Atti,}{
  address={Department of Physics, University of Perugia and
    Istituto Nazionale di Fisica Nucleare, Sezione di Perugia,
    Via A. Pascoli, I-06123, Italy}}
\author{C. B. Mezzetti}{
  address={Department of Physics, University of Perugia and
    Istituto Nazionale di Fisica Nucleare, Sezione di Perugia,
    Via A. Pascoli, I-06123, Italy}}
\author{V. Palli}{
  address={Department of Physics, University of Perugia and
    Istituto Nazionale di Fisica Nucleare, Sezione di Perugia,
    Via A. Pascoli, I-06123, Italy}}
\author{S. Scopetta}{
  address={Department of Physics, University of Perugia and
    Istituto Nazionale di Fisica Nucleare, Sezione di Perugia,
    Via A. Pascoli, I-06123, Italy}}
\author{{ }L. P. Kaptari}{address={ Bogoliubov Laboratory of
    Theoretical  Physics,141980, JINR,  Dubna, Russia}}
\author{H. Morita}{
  address={Sapporo Gakuin University, Bunkyo-dai 11, Ebetsu 069,
    Hokkaido, Japan}}
\begin{abstract}
The effects of short range correlations in lepton and hadron
scattering off nuclei at medium and high energies are discussed.
\end{abstract}
\maketitle
\section{Introduction}
There is at present  an uprise of interest in the longstanding
problem   concerning the role played by Nucleon-Nucleon (NN) short range
correlations (SRC) in medium and high energy scattering off nuclei
thanks particularly to the results of a series of dedicated
experiments providing quantitative information on proton-neutron
(p-n) and proton-proton (p-p) SRC in nuclei \cite{eli}. In  view
of these results,  as well as of previous demonstrations
of  the inadequacy of the mean field picture of nuclei to
describe  the high momentum behavior of the nuclear wave
functions (see e.g. \cite{cps}), it is timely to carefully analyze
the effects of SRC  in various scattering processes,
paying particular attention  to those competing effects, particularly
final state interaction (FSI) effects, which make the study of SRC
not easy task. In what follows the results of several calculations
performed at the University of Perugia along this line  will be presented.
\begin{figure}
\centerline{
\includegraphics[height=6.5cm,width=5.5cm]{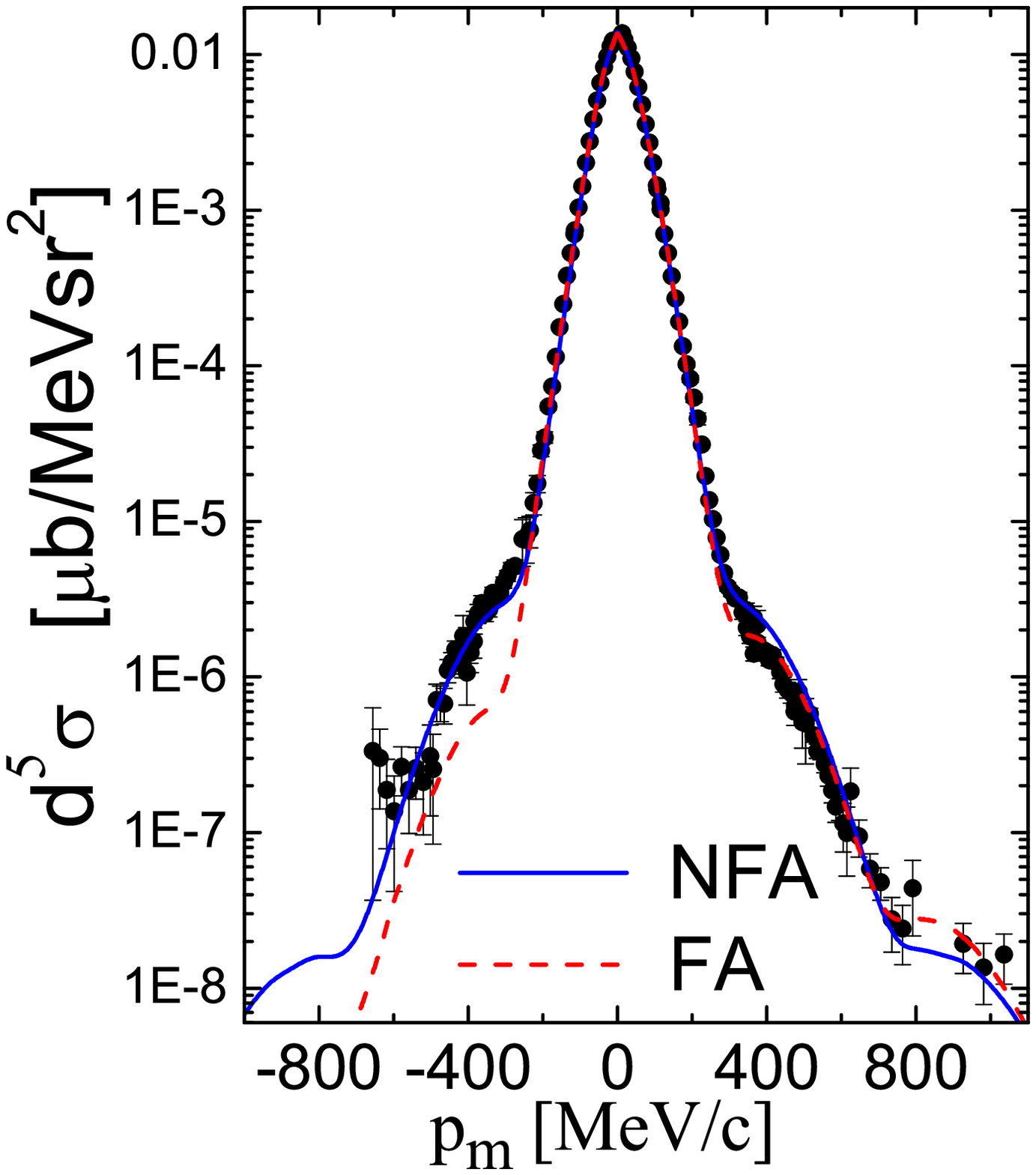}
\hspace{-1.0cm}
\includegraphics[height=6.6cm,width=5.7cm]{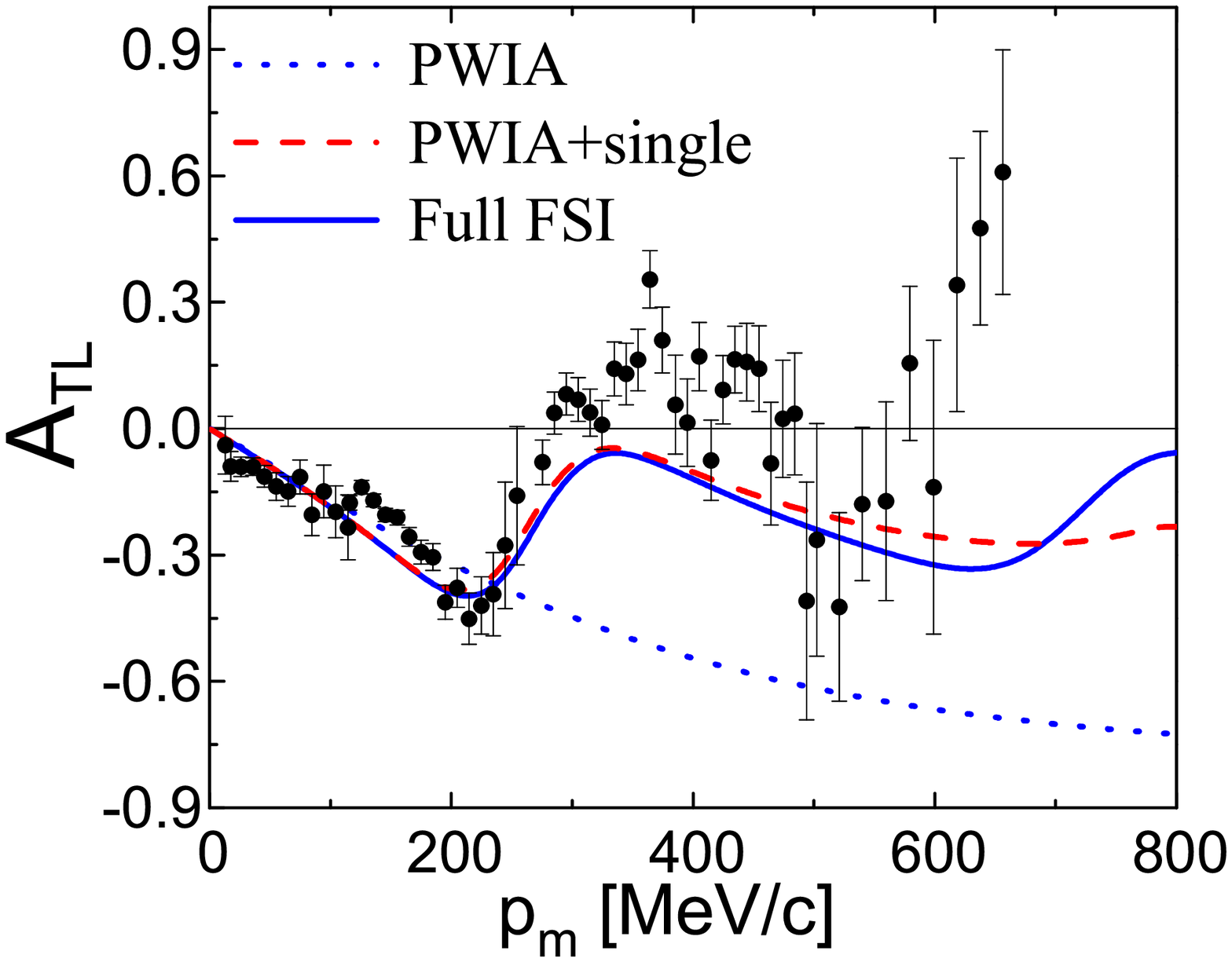}
\hspace{-0.8cm}
\includegraphics[height=6.5cm,width=5.7cm]{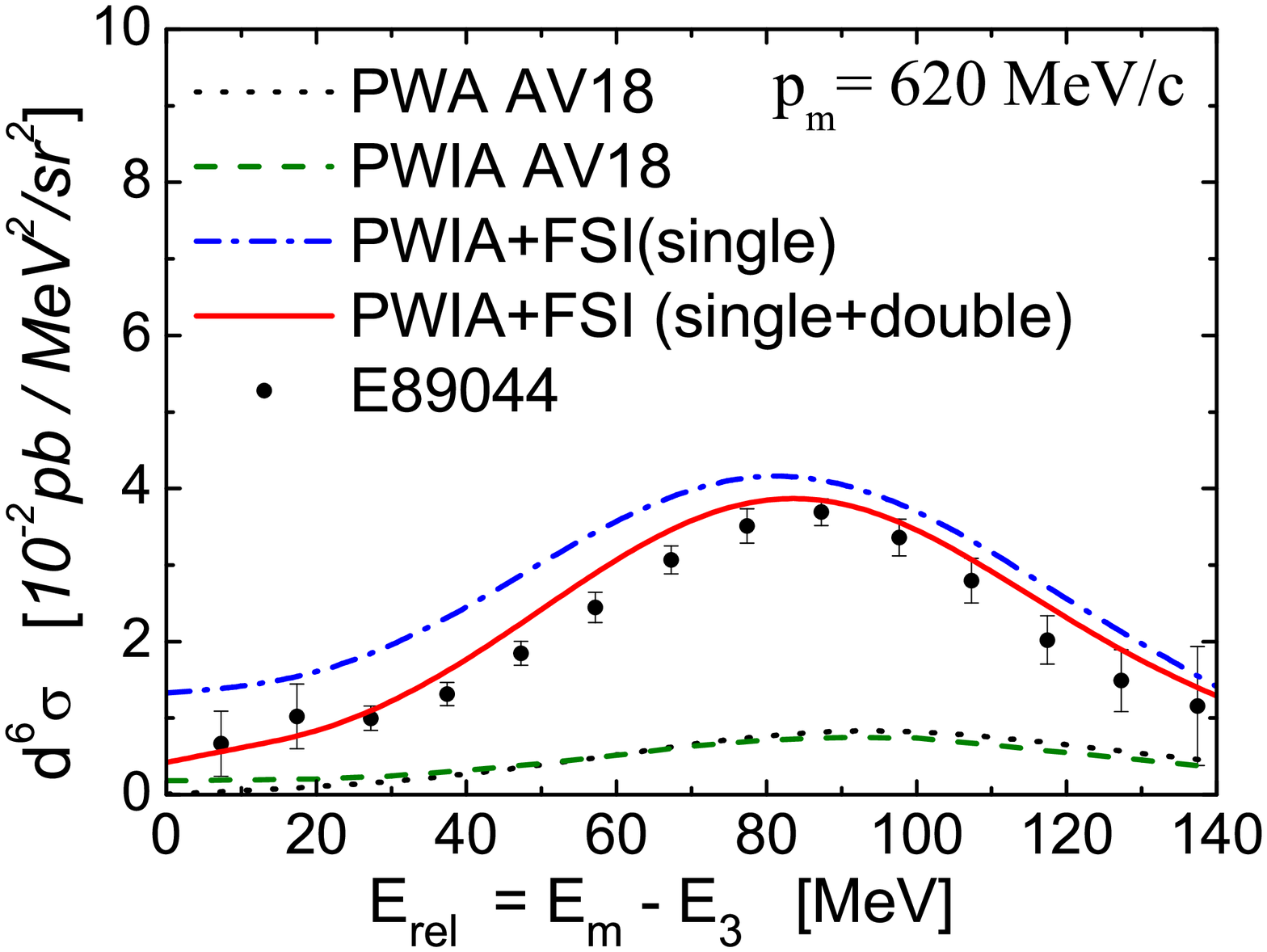}
}
 \caption{LEFT: the  five-fold cross section $d^5\sigma \equiv d^5\sigma
 \times{(dE'd\Omega'd\Omega_{p_1})}^{-1}$
 of the 2bbu channel
$^3He(e,e^\prime p)^2H$  calculated in  GEA within
the non factorized (NFA) and factorized (FA) approaches {\it vs}
 the missing momentum $p_m$.
\,\, CENTER: the $A_{TL}$
asymmetry. \,\, RIGHT: the six-fold cross section $d^6\sigma \equiv d^6\sigma
 \times{(dE'd\Omega'd\Omega_{p_1}dE_m)}^{-1}$ of the 3bbu channel
  plotted, at fixed value of $p_m$, {\it vs}
the excitation energy $E_{rel}$ of the two-nucleon system in the continuum.
   Experimental data from  \cite{rvachev,E89044} (After Ref.
\cite{nofac})}
      \label{Fig1}
\end{figure}

\section{The ${\bf A(e,e'p)B}$ process in few-nucleon systems}
Exclusive  lepton scattering off nuclei $A(e,e'p)B$ in the quasi elastic
region, plays a relevant role in nowadays hadronic physics, since
it can provide information on: i) SRC; ii) the details of the mechanism
of propagation of hadronic states in the medium; iii)  QCD
predictions like, e.g.,  color transparency effects. At medium and
high energies the propagation of a struck hadron in the medium  is
usually treated within the Glauber multiple scattering approach
(GA),  which has been applied with great success to  hadron
scattering off nuclear targets. However, when the hadron is
created inside the nucleus various improvements of the original GA
have been advocated; worth being mentioned is the  generalized eikonal
approximation (GEA) \cite{Misak05} where,  unlike GA,  the
excitation energy of the  $A-1$ system is partly taken into
account.  The GEA has recently  been applied  to a systematic
calculation of the
 two-body (2bbu)  and three-body  (3bbu)  break up channels of the
 $^3He(e,e'p)X$ reaction \cite{CiofiRev,rocco}
 using realistic three-body  wave functions \cite{Kiev,VMC}
 and interactions \cite{AV18}.
In GEA the final state wave function has  the following form (spin and isospin
variables are omitted for ease of presentation)
\be \Psi_f({\bf r}_1,\,.\,.\,.{\bf r}_A)={\hat{\cal A}} S_{GEA}({\bf
r}_1,\,.\,.\,.{\bf r}_A)e^{-i {\bf p}_1{\bf r}_1} e^{-i {\bf P}_{A-1} {\bf
R}_{A-1}} \Phi_{A-1}({\bf r}_2,\,.\,.\,.{\bf r}_A) \label{states} \ee
 \noindent where
 $ {S}_{GEA}=\sum \limits_{n=1}^{A-1}
{S}_{GEA}^{(n)}$   takes care of the FSI;  here, $n$ denotes  the
order of multiple scattering, with the  single scattering term
($n$=1) given by ${S}_{GEA}^{(1)}({\bf
r}_1,\,.\,.\,.{\bf r}_A)=1-\sum\limits_{i=2}^A
\theta(z_i-z_1){\rm e}^{i\Delta_z (z_i-z_1)} \Gamma (\b
{b}_1-\b{b}_i)$,  where $\Gamma({\bf b})
={\sigma_{NN}^{tot}(1-i\alpha_{NN})} \cdot exp\,[-{\bf
b}^2/2b_0^2]/[4\pi b_0^2]$ is the usual Glauber profile function
and $\Delta_z = ({q_0}/{|{\b q}|}) E_m$, $E_m$ being the removal
energy related to the excitation energy of $A-1$. Note that
when $\Delta_z=0$, the usual GA is recovered.
A fully non-factorized calculation (NFA) of the 2bbu and 3bbu channels
has been recently performed \cite{nofac} within a fully
parameter free GEA approach. The results are shown
in  Fig. \ref{Fig1}. It can be seen that in the 2bbu channel the
factorized approximation (FA) is a poor one  in the left region
($\phi = 0$, $\phi$ being the angle between the scattering and reaction
planes) of the missing momentum, unlike what happens in the right region
($\phi =\pi$). The quantitative disagreement between theory and experiment
is visualized in terms of the left-right asymmetry  $A_{TL}
=[d\sigma(\phi=0^o)-d\sigma(\phi=180^o)]/[d\sigma(\phi=0^o)+d\sigma(\phi=180^o)]$.
 As for the 3bbu channel shown in the same Figure, the
differences between the FA and NFA amount at most at $10-15 \%$.
\begin{figure}
\centerline{\includegraphics[width=7.0cm,height=7.0cm]{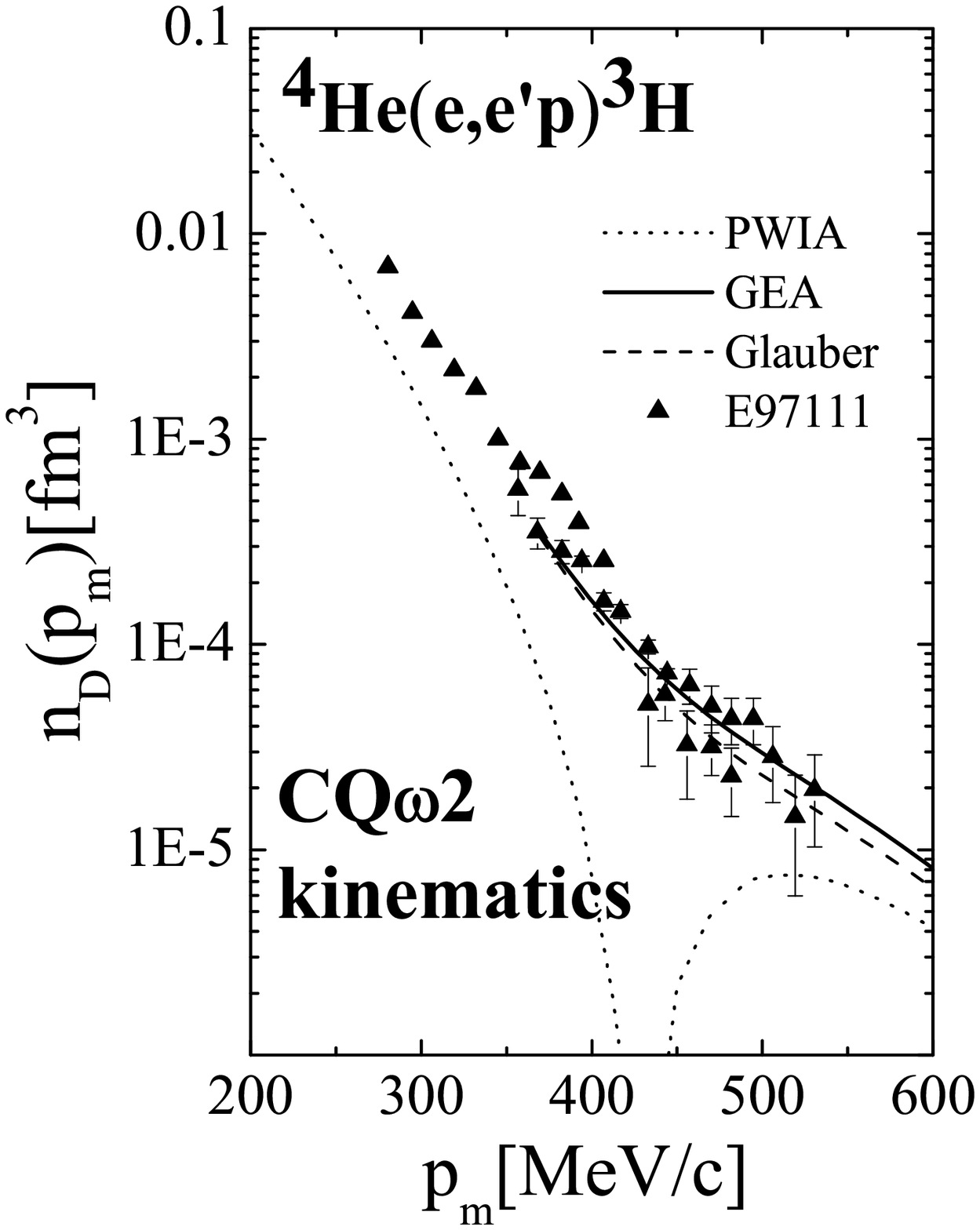}
\hspace{-0.8cm}\includegraphics[width=6.4cm,height=7.0cm]{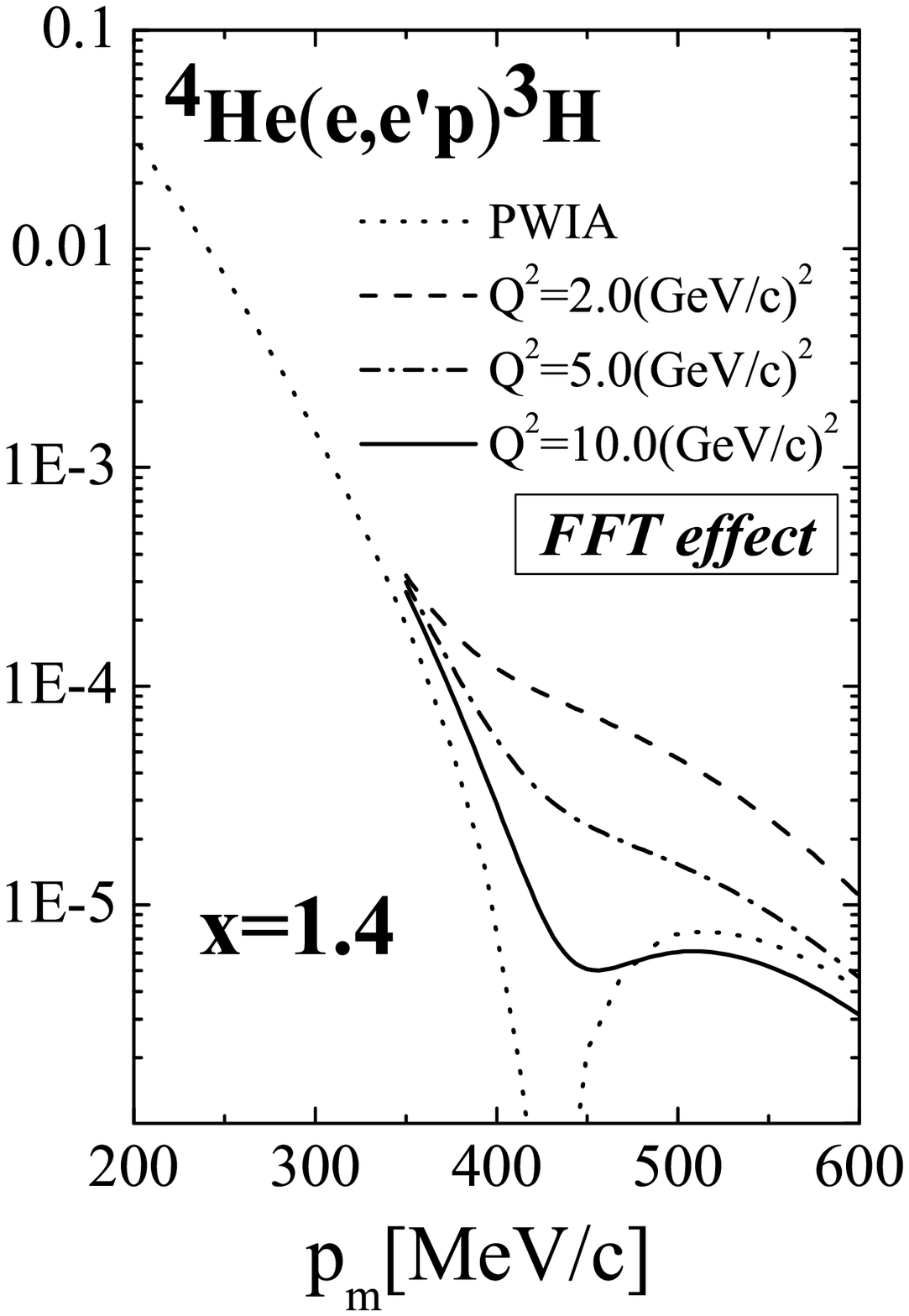}}
\caption{LEFT:\,\, the reduced cross section ($n_D(\mathbf{p}_m)=
({d^5 \sigma}/ d \Omega ' d {E'}d \Omega_{p_1}) \times ({\mathcal
K} \sigma_{ep})^{-1}$)  of the process $^4He(e,e'p)^3H$ at
perpendicular
 kinematics and  $x \simeq 1.8$. The solid line shows the results within GEA, whereas
 the dashed curve corresponds to the conventional GA.
  Preliminary  data from \cite{E97111}.
RIGHT:\,\, the reduced cross section  at perpendicular
 kinematics for various values of $Q^2$ and  $x \simeq 1.4$, calculated taking
  FFT effects \cite{Braun} into
 account. Four-body wave functions from  \cite{ATMS2}. Experimental data from
 \cite{reitz}.
 (After Refs. \cite{CKM}, \cite{FFTHiko} and \cite{ckmfewbody})}
  \label{Fig2}
\end{figure}
The results for the  process  $^4He(e,e^\prime p)^3He$ are shown in
Fig. \ref{Fig2}. It can be seen that: i) the dip predicted by the PWIA
is completely  filled up  by the FSI; ii)  like the $^3He$ case, the
difference between GA and  GEA  is not very large; iii) an overall
satisfactory agreement between theory and experiment is obtained.
In Fig. \ref{Fig2} the results which take into account the finite formation
time (FFT) of the proton are also shown. In \cite{Braun} FFT effects have
been implemented by explicitly considering the  dependence
of the NN scattering amplitude upon nucleon  virtuality, leading to a replacement of
$\theta(z_i-z_1)$, appearing in the Glauber profile,  with
$\textit{J}(z_i-z_1)= \theta(z_i-z_1)\left( 1-\exp [-(z_i-z_1)/{\it l}(Q^2)] \right)$
  where  ${\it l}(Q^2)={Q^2}/(x m_N\,M^2)$; here
  $x$ is the  Bjorken scaling  variable and the quantity $l(Q^2)$ plays
the role of the proton
 formation length, i.e. the length of the trajectory that the
knocked out proton runs until it returns to its asymptotic form;
the quantity $M^2$ is $M^2 = {m^*}^2 - m_N^2$ and
 since the formation length grows linearly with
$Q^2$, at higher $Q^2$ the strength of the Glauber-type FSI is
reduced by the damping factor $( 1-\exp[-(z_i-z_1)/{\it
l}(Q^2)])$.
It can be seen that at the JLAB
kinematics  ($Q^2=1.78\,(GeV/c)^2$, $x\sim1.8$) FFT effects, as expected,
 are too small to be detected, but they can unambiguously be observed
in the region $5 \leq Q^2 \leq 10\, (GeV/c)^2$ and   $x$ =1.4.

\section{${\bf A(e,e')X}$ and  ${\bf A(e,e'p)B}$ processes off complex nuclei}

Recently \cite{acmprc2005,fizika,acmprl2008} the ground state properties
(energy, density and momentum distributions) of closed shell and sub-shell
nuclei have been obtained by a cluster expansion approach based
upon a correlated wave function
\begin{figure}
\centerline{
\includegraphics[height=5.0cm,width=6.4cm]{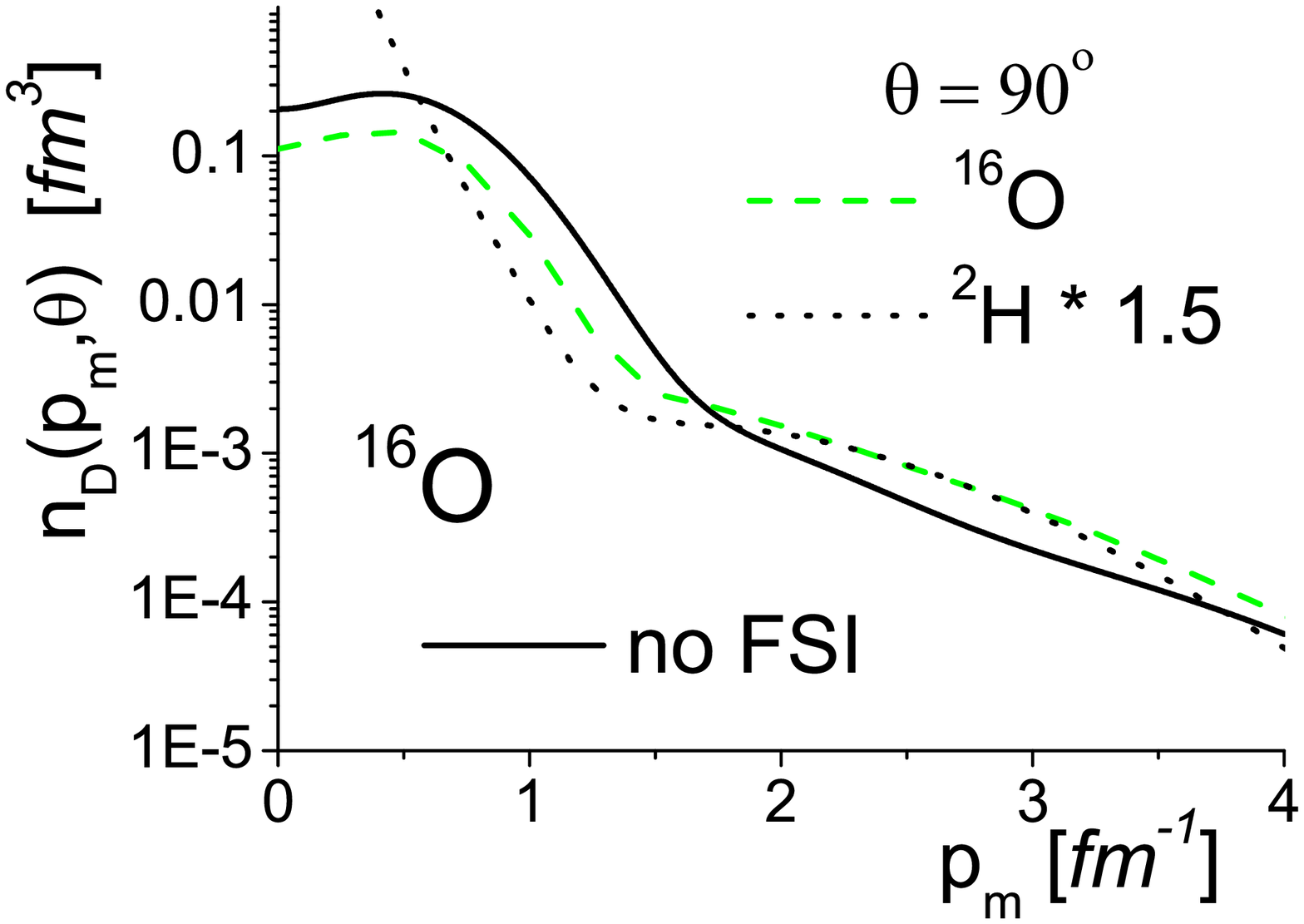}
\hspace{-1.0cm}
\includegraphics[height=5.0cm,width=5.3cm]{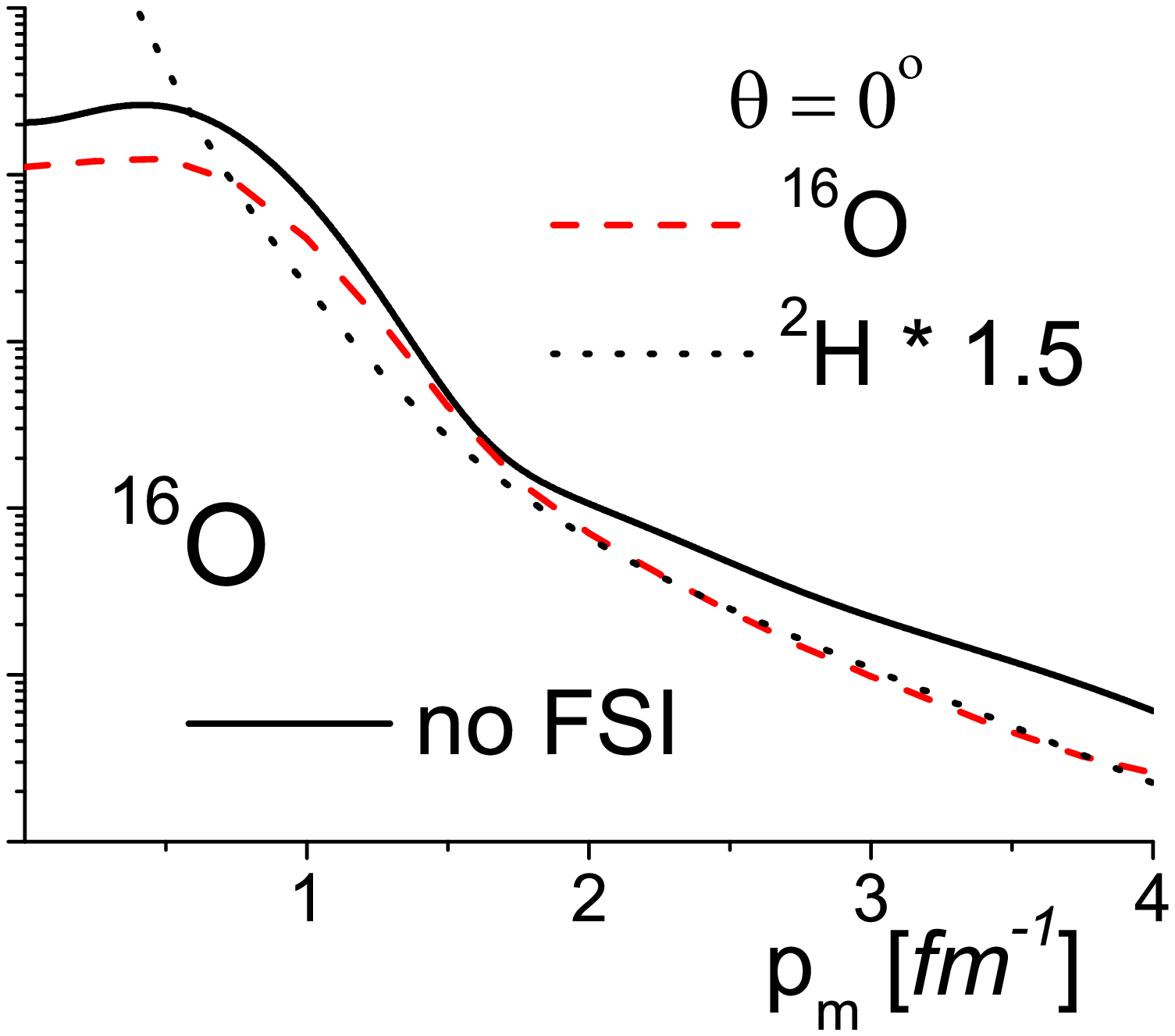}
\hspace{-1.0cm}
\includegraphics[height=5.0cm,width=5.3cm]{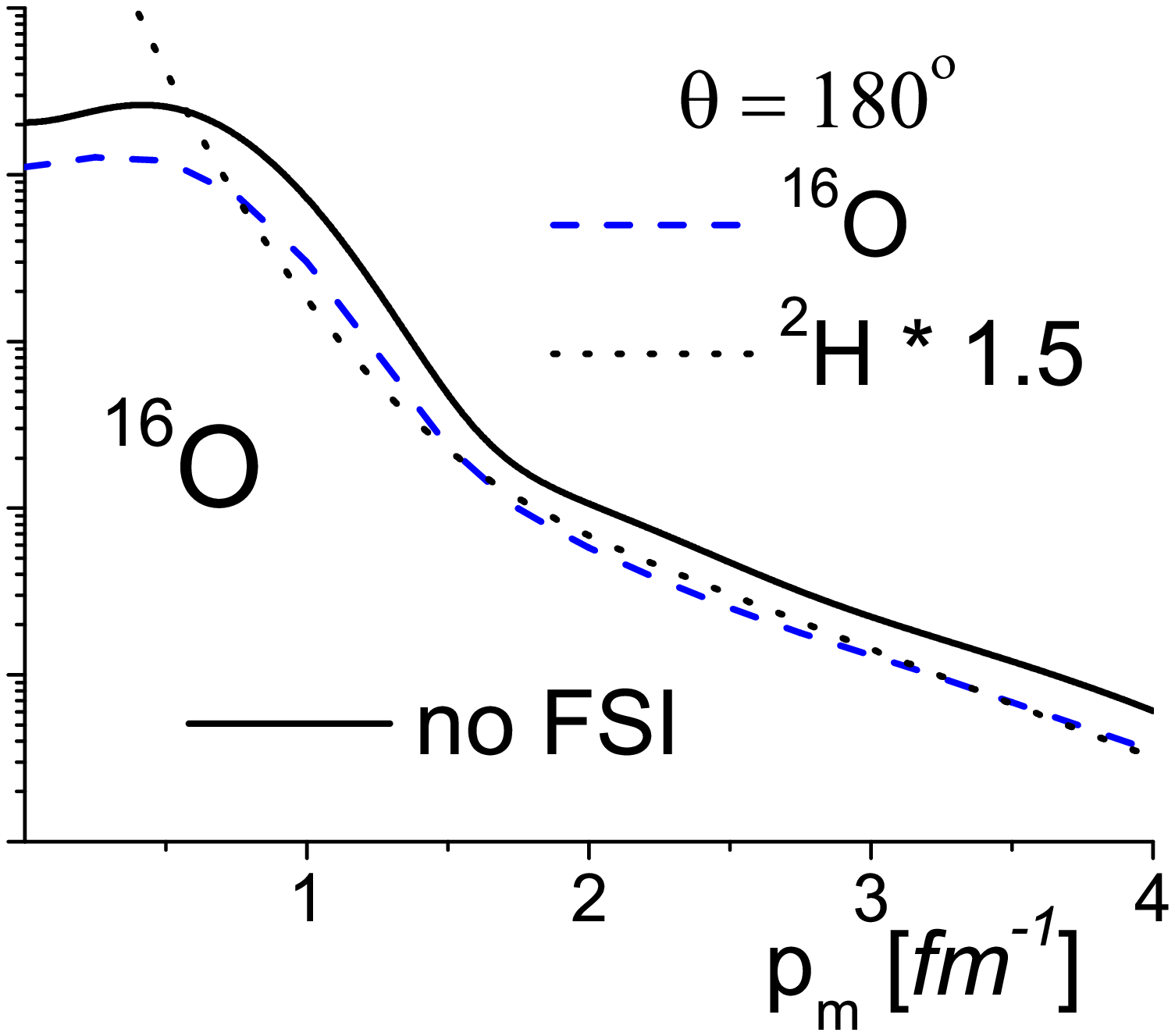}}
  \caption{The distorted missing momentum distributions of $^{16}O$ at
  various angles (dashes) compared with the rescaled deuteron distorted
  momentum distribution at the same angles (dots). The full line represents
  the undistorted momentum distribution. Wave functions from
  Ref. \cite{acmprc2005}. (After Ref. \cite{distorted})}
  \label{Fig3}
\end{figure}
%
\begin{figure}
\centerline{\epsfysize=7.0cm\epsfbox{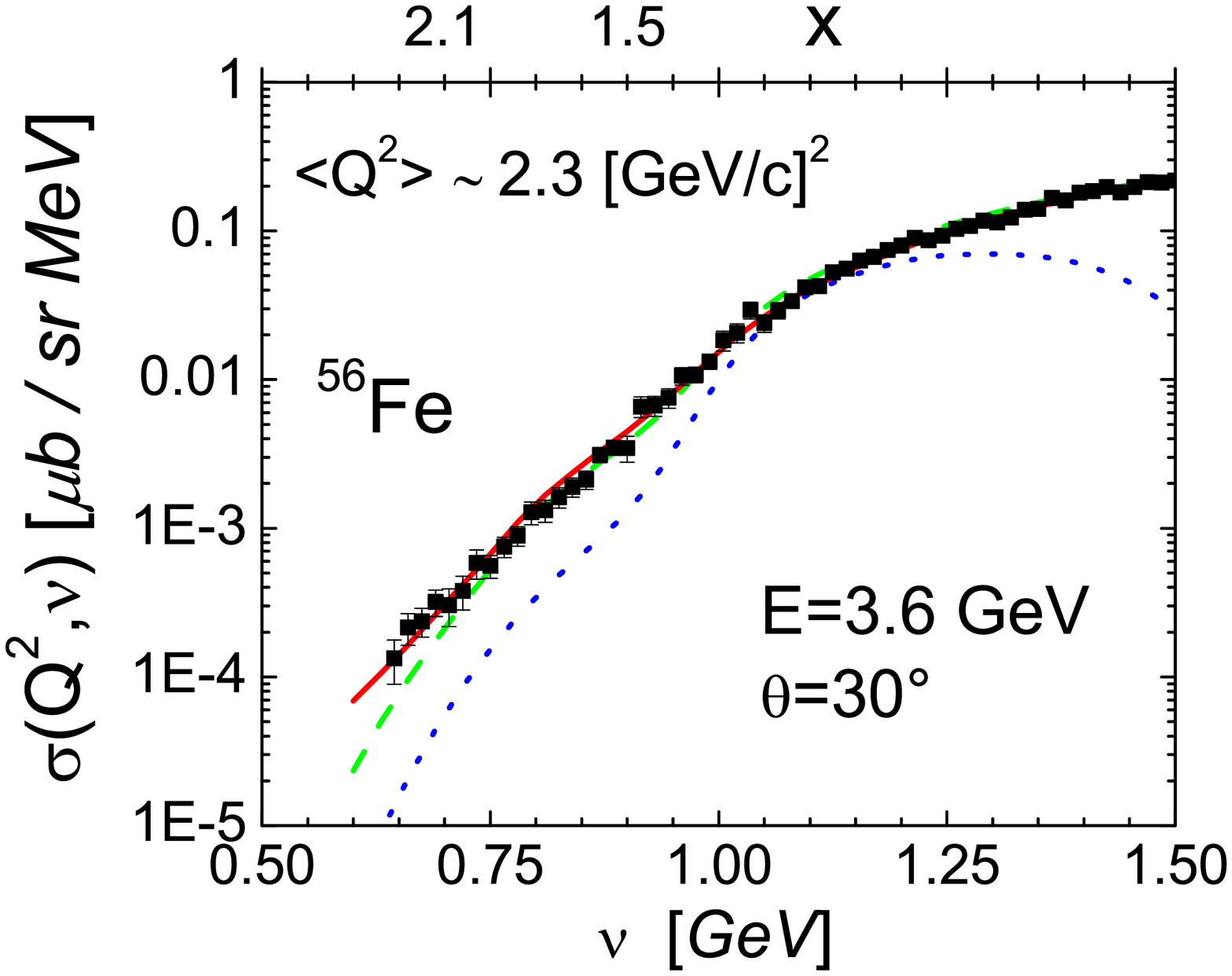}
\hspace{-1.0cm}\epsfysize=7.0cm\epsfbox{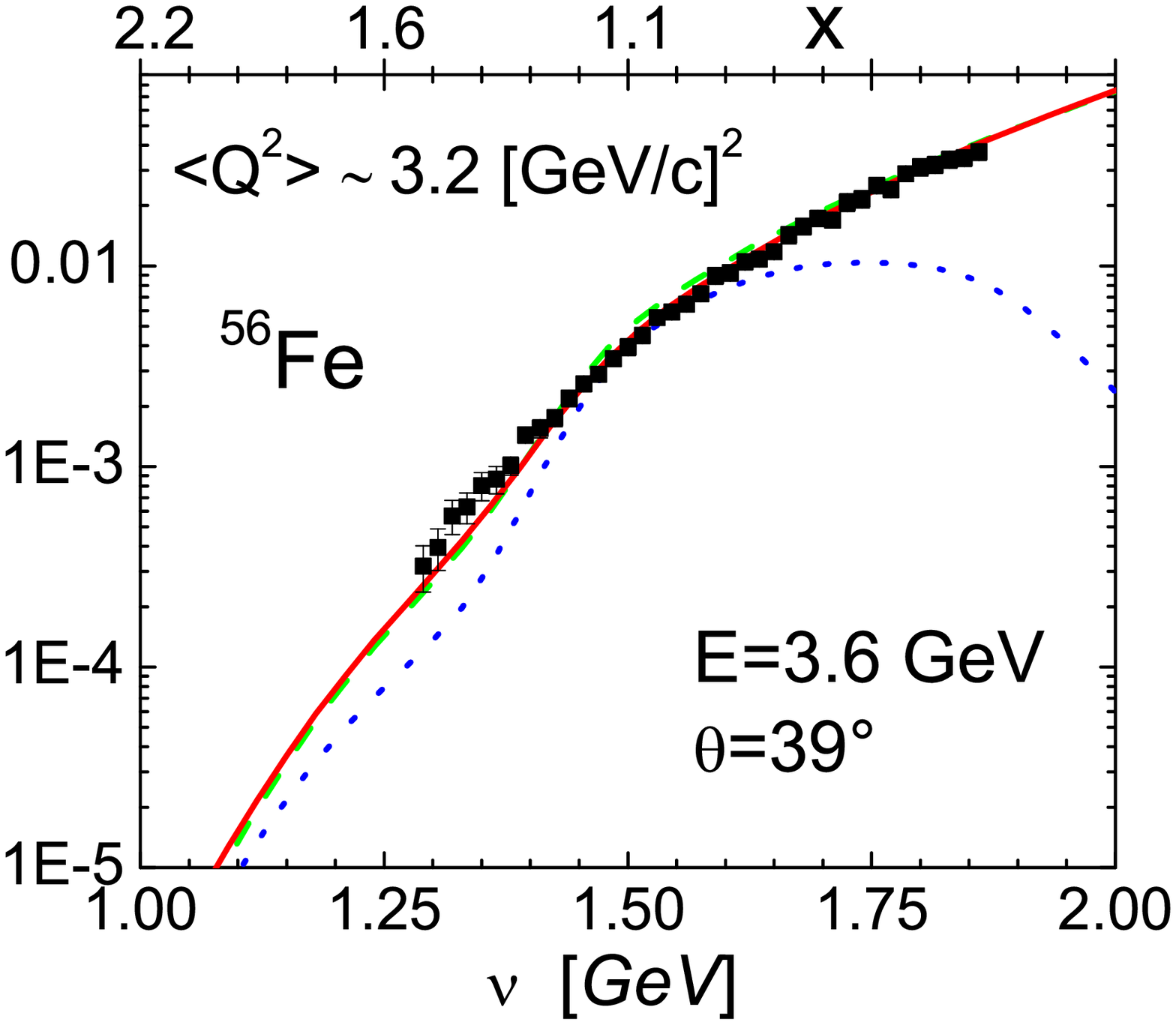}}
\caption{The experimental inclusive cross section $^{56}Fe(e,e')X$
\cite{slac} compared with theoretical calculations which include
SRC \cite{ciosim} and FSI \cite{ciosim1}.  {\bf \textit{Dotted
lines}}: PWIA;
 {\bf \textit{dashed lines}}: PWIA + FSI of the correlated struck nucleon with
 the correlated partner;
{\bf \textit{full lines}}: the same as {\bf\textit{dashed lines}}
plus the FSI of the shell model struck nucleon with the mean field
of the $(A-1)$ spectator. (After Ref. \cite{clach})} \label{Fig4}
\end{figure}
%
\begin{figure}
\epsfxsize=8.2cm\epsfbox{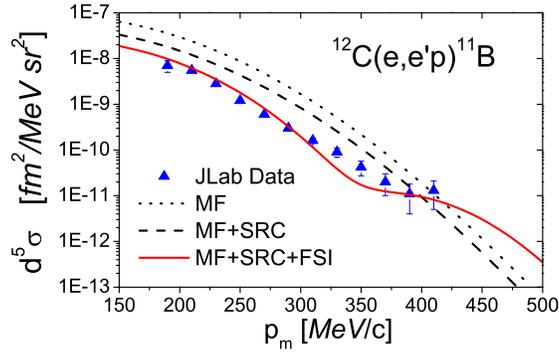} \caption{The preliminary
Jlab data \cite{E01-015} on of the exclusive reaction
$^{12}C(e,e'p)^{11}B$ compared with Mean Field ($MF$) predictions
({\it dots}) and with the predictions of $MF$ + $SRC$ (${\it
dashes}$) and $MF$+$SRC$ + $FSI$ (${\it full}$); $d^5\sigma \equiv
d^5\sigma \times(d\Omega'd\Omega_{p_1}dE_{p_1})^{-1}$. (After Ref.
\cite{acmexclu})} \label{Fig5}
\end{figure}
\be
\psi_o({\bf r}_1,{\bf r}_2,...,{\bf r}_A)\,
=\,{{\hat{F}}({\bf r}_1,{\bf r}_2,...,{\bf r}_A)}
\,\phi_o({\bf r}_1,{\bf r}_2,...,{\bf r}_A)
\ee
 where $\phi_o$ is a \textit{mean-field} wave function and
\be
{\hat{F}}\,=\,\hat{S}\,\prod_{i<j}\,\hat{f}_{ij}\,=\,\hat{S}\,
\prod_{i<j}\,\sum_{n=1}^{N}\,f^{(n)}(r_{ij})\,
\hat{\mathcal{O}}^{(n)}_{ij}
\ee
is a symmetrized \textit{correlation}
operator generating NN correlations according to the modern forms
 of the two-nucleon interaction, i.e.
\cite{AV18}
$\hat{V(i,j)}\,=\,\sum_{n=1}^{N}\,v^{(n)}(r_{ij})\,\mh^{(n)}_{ij}$
where  the state-dependent operator is
$\mh^{(n)}_{ij}\,=\,\left[1\,,\,{\Bf\sigma}_i\cdot{\Bf\sigma}_j\,
        ,\,\hat{S}_{ij}\,,\,\left({\bf S}\cdot{\bf L}\right)_{ij}\,,
        \,L^2\,,\,L^2{\bf\sigma}_i\cdot{\Bf\sigma}_j\,,\,
      \left({\bf S}\cdot{\bf L}\right)^2_{ij}\,,..
        \right]\,\otimes\,\left[1\,,\,{\bf \tau}_i\cdot{\bf \tau}_j\right]$.
 The ground state properties have been obtained by cluster expanding
 the appropriate density matrices. Using the wave function parameters which
  minimize the ground
 state energy and GEA, a realistic approach to SRC and FSI is achieved.
  In Fig. \ref{Fig3} the distorted momentum distributions
   for $^{16}O$ are shown and compared with the distorted momentum distribution of
   the deuteron, finding, at high values of the missing momentum,  an amazing
   similarity between them, as in the case of the undistorted momentum distributions
   \cite{ciosim}.
The inclusive cross section calculated by separately considering FSI in the correlated
pair and in the $(A-1)$ mean field \cite{ciosim,ciosim1} are presented in
Fig. \ref{Fig4}, which clearly demonstrates the relevant role played by FSI in the SRC pair.
A comparison between preliminary Jlab data on the exclusive process
 $^{12}C(e,e'p)^{11}B$ and theoretical calculations based upon
correlated wave functions and Glauber FSI, are presented in Fig. \ref{Fig5}.

\section{The tensor force and  ${\bf pn}$/${\bf pp}$
correlations}
The  role of the tensor force in producing a
substantial difference between $pn$ and $pp$ two-nucleon momentum
distributions in few-body systems and light nuclei ($A \leq 8$),
has been recently demonstrated \cite{schiavilla,schiavilla1}
 using state-of-the-art  realistic nuclear wave functions obtained within the
 Variational
Monte Carlo (VMC) approach \cite{VMC}. The same conclusion for complex nuclei
($12 \leq A \leq 40$) has been reached in \cite{acmprl2008} within the approach of
Ref. \cite{acmprc2005}.
The two-body Momentum Distribution is (${\bf k}_{rel}=({\bf k}_1-{\bf k}_2)/2$,
 ${\bf K}_{CM}=({\bf k}_1+{\bf k}_2)$)
\beqy
n({\bf k}_{rel},{\bf K}_{CM})&=&\frac{1}{(2\pi)^6}\int
d{\bf r}d{\bf r}^\prime
d{\bf R}d{\bf R}^\prime\,e^{-i\,{\bf K}_{CM}\cdot({\bf R}-{\bf R}^\prime)}
e^{-i\,{\bf k}_{rel}\cdot({\bf r}-{\bf r}^\prime)}
\rho^{(2)}({\bf r},{\bf r}^\prime;{\bf R},{\bf R}^\prime)\nonumber
\eeqy
\begin{figure}
\epsfysize=4.6cm\epsfbox{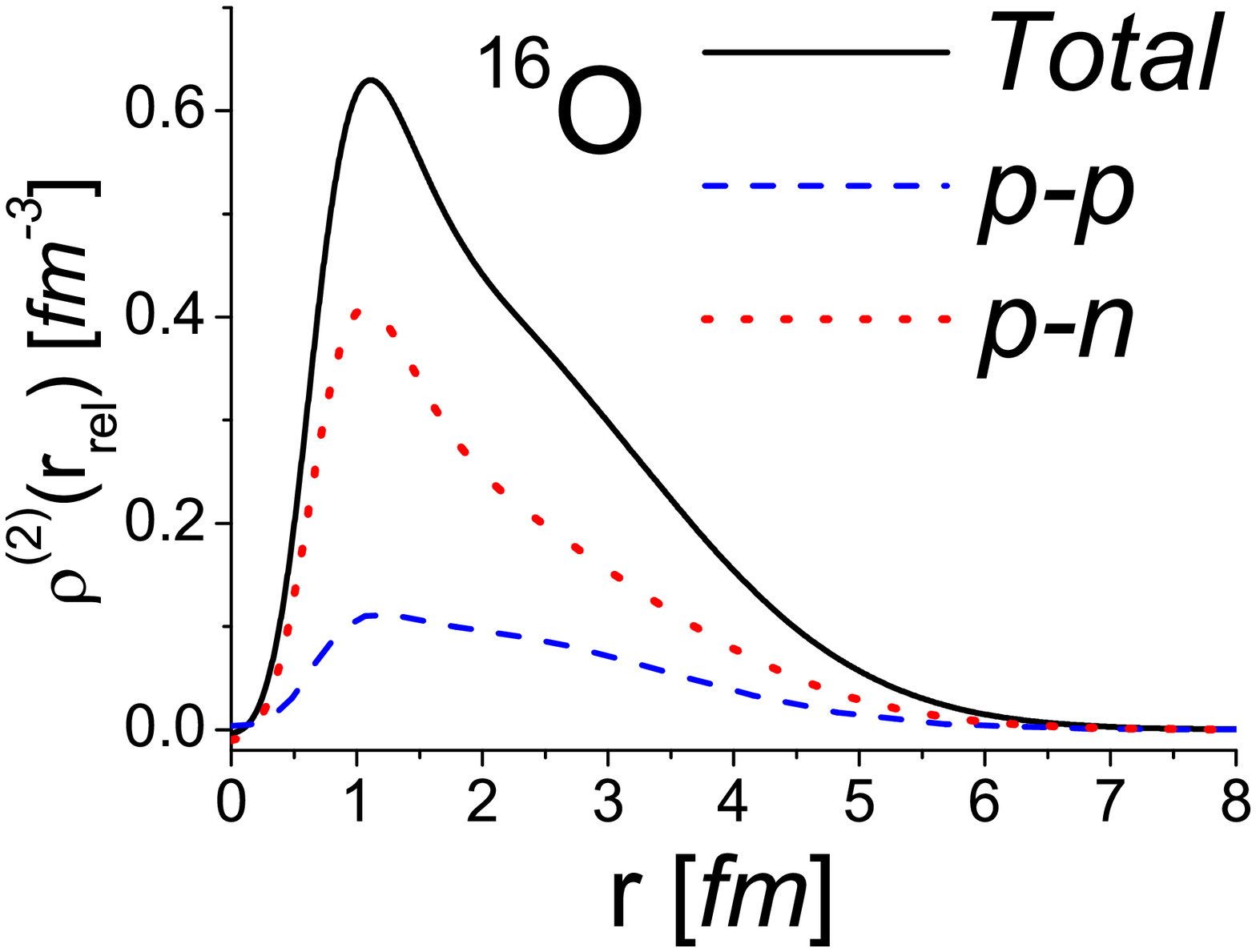}\hspace{-0.8cm}
\epsfysize=4.6cm\epsfbox{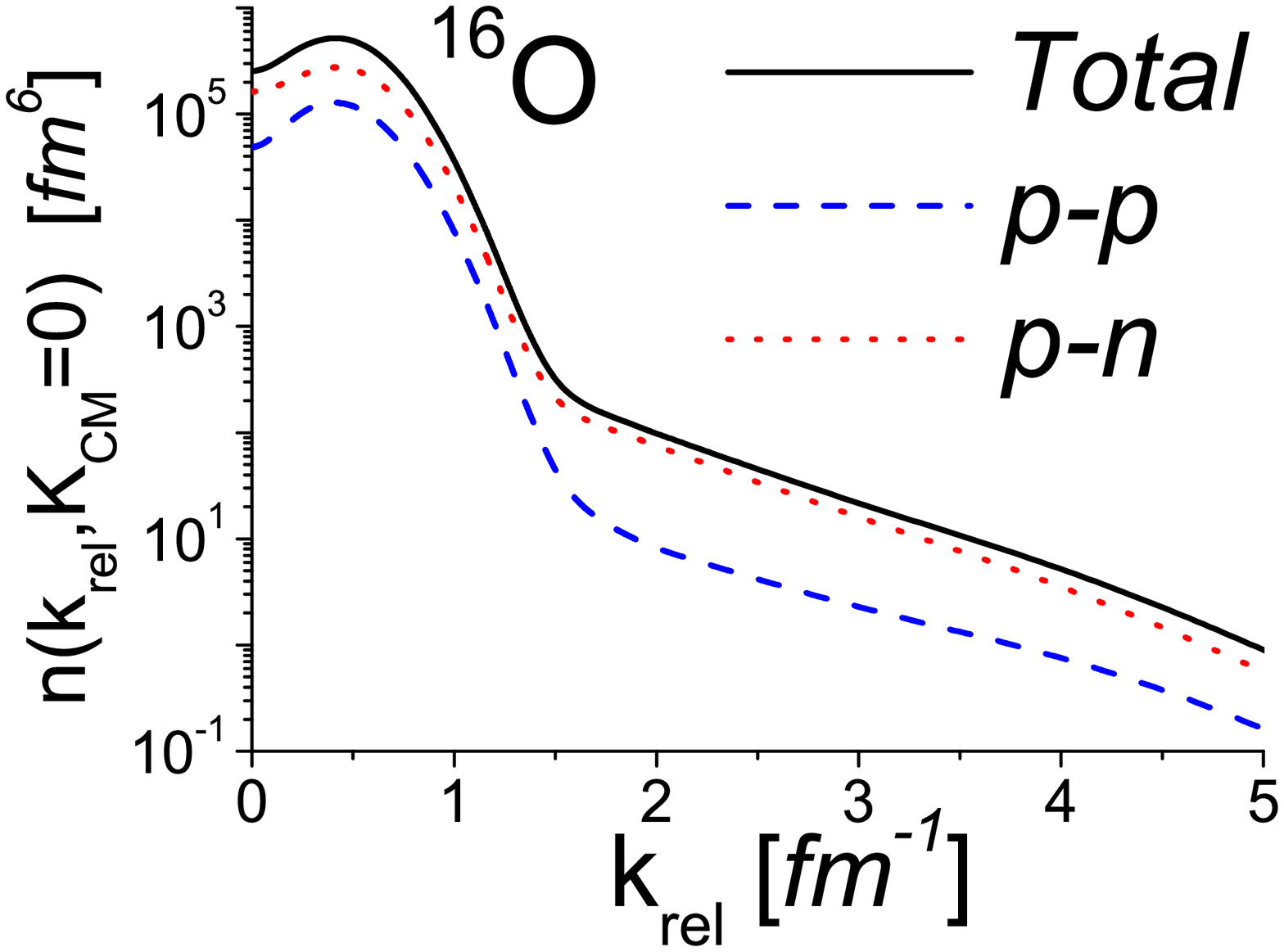}\hspace{-1.1cm}
\epsfysize=4.6cm\epsfbox{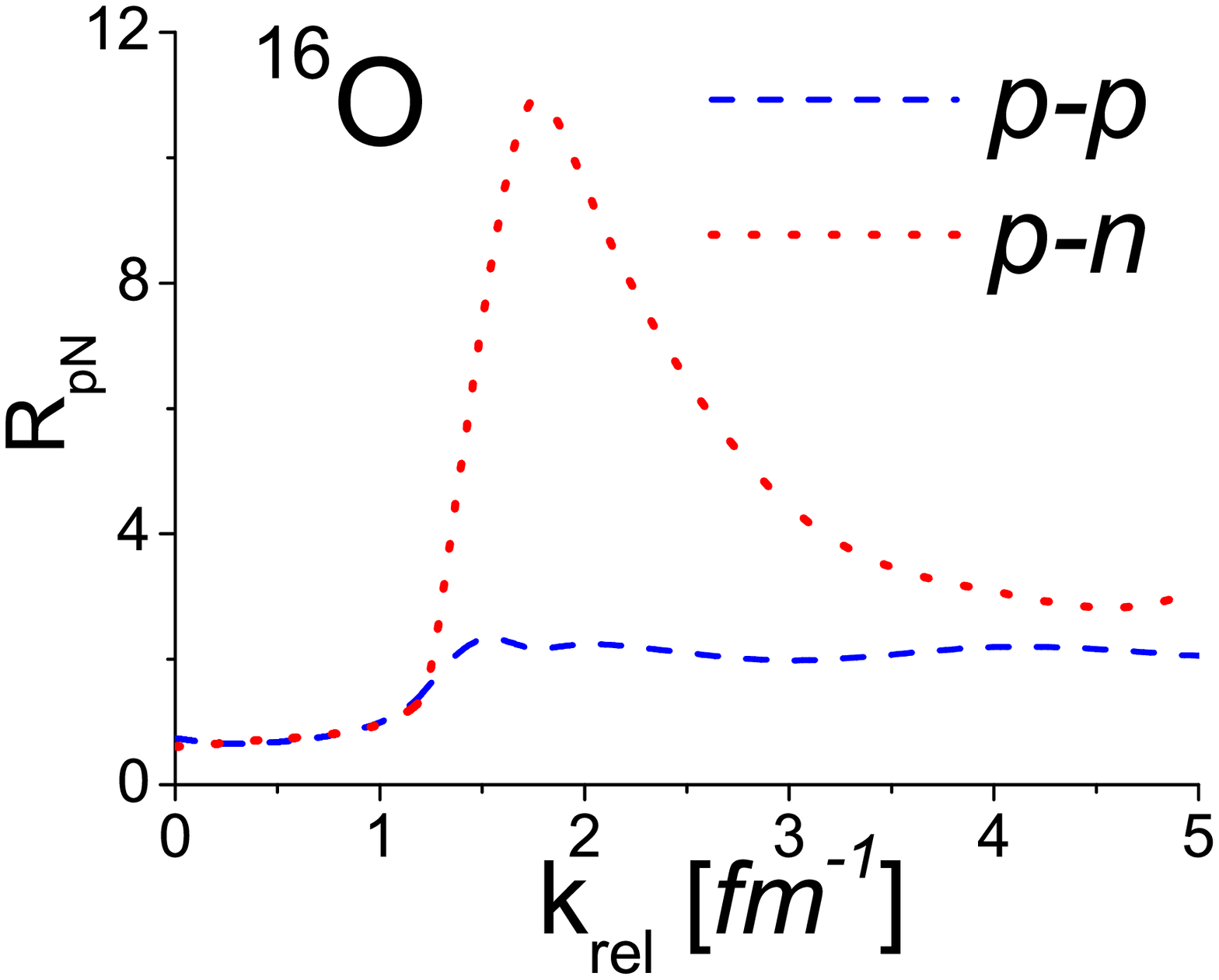} \caption{LEFT: the
$^{16}O$  relative (integrated over ${\bf R}$)
 diagonal two-body density distribution.\,
CENTER: the  two-body momentum distribution for ${\bf K}_{CM}=0$. RIGHT: the
 ratio
$R_{pN}=n_{pN}(k_{rel},0)\,/\,n^{central}_{pN}(k_{rel},0)$. (After Refs.
\cite{acmprl2008,ACM1}}
\label{Fig6}
\end{figure}
\noindent where $\rho^{(2)}({\bf r},{\bf r}^\prime; {\bf R},{\bf R}^\prime)$ is the
non-diagonal two-body density matrix. The relative (integrated over $\bf R$)
density matrix $\rho^{(2)}({\bf r}_{rel})$,
the quantity $n({\bf k}_{rel},{\bf K}_{CM}=0)$, describing
back-to-back nucleons, and the effects from the
tensor force on $n_{pN}({\bf k}_{rel},{\bf K}_{CM}=0)$, characterized by the ratio
$R_{pN}=n_{pN}(k_{rel},0)\,/\,n^{central}_{pN}(k_{rel},0)$, are shown in Fig.
\begin{table}[!htp]
    \caption{The $pp$ and $pn$ percentage probability (Eq. (\ref{percent}))
      evaluated in the momentum range shown in square brackets.}
    \begin{tabular*}{0.45\textwidth}{@{\extracolsep{\fill}}c||c|c||c|c}
      & $P_{pp}$ ($\%$) & $P_{pn}$ ($\%$) & $P_{pp}$ ($\%$) & $P_{pn}$ ($\%$)\\
      $A\,\,\,$  & $[0\,,\,\infty]\,\,$ & $[0\,,\,\infty]\,\,$ & $[1.5\,,\,3.0]\,\,$
                                                      & $[1.5\,,\,3.0]\,\,$ \\\hline
      $4\,\,\,$  & $19.7$ & $81.3$ & $2.9$  & $97.1$\\
      $12\,\,\,$ & $30.6$ & $69.4$ & $13.3$ & $86.7$\\
      $16\,\,\,$ & $29.5$ & $70.5$ & $10.8$ & $89.2$\\
      $40\,\,\,$ & $31.0$ & $69.0$ & $24.0$ & $76.0$
    \end{tabular*}\label{Tab1}
\end{table}
\ref{Fig6}. In ref. \cite{acmprl2008} the percentage
probability to find in the nucleus a $pN$ pair defined by
\be
P_{pN}\,=\,\frac{\int^b_a dk_{rel}\,k^2_{rel}\,n_{pN}({k}_{rel},0)}{
\int^b_a dk_{rel}\,k^2_{rel}\,\left(\,n_{pp}({k}_{rel},0)
\,+\,\,n_{pn}({k}_{rel},0\right)}\,.
\label{percent}
\ee
has been calculated. The results are shown in Table \ref{Tab1}. It can be seen
that, in agreement with the result of Ref. \cite{schiavilla}, when the integration runs
over the whole range of $k_{rel}$,  $P_{pN}$ is proportional to the percentage of
$pN$ pairs, whereas the integration over the correlation region leads to a
percentage of $pn$ pairs much larger  than that of the $pp$ pairs, which is clear
consequence of the effects of the tensor  force acting
\begin{figure}
\centerline{
\includegraphics[width=8.9cm,height=7.5cm]{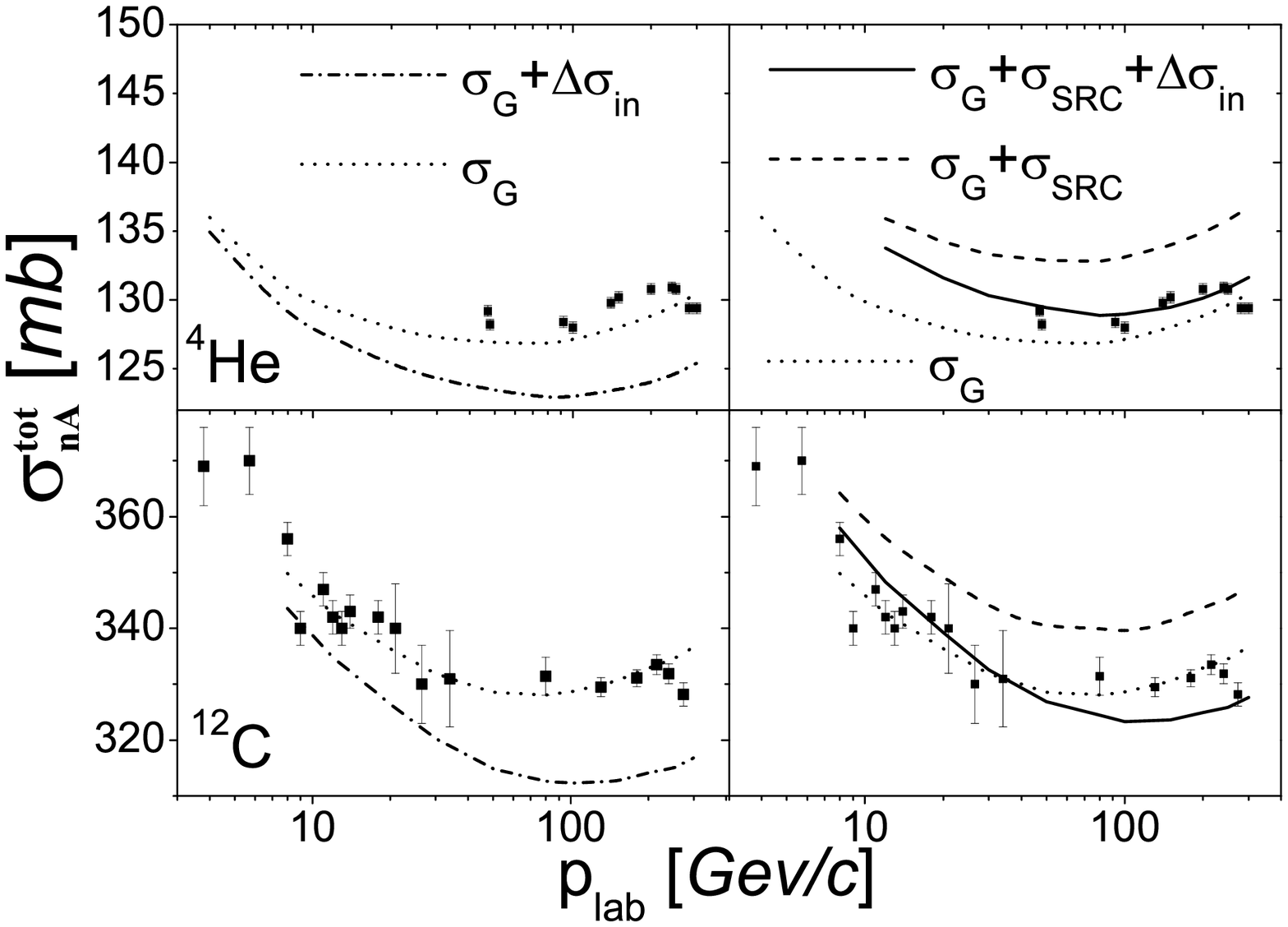}
\hspace{-1.2cm}
\includegraphics[width=8.3cm,height=7.5cm]{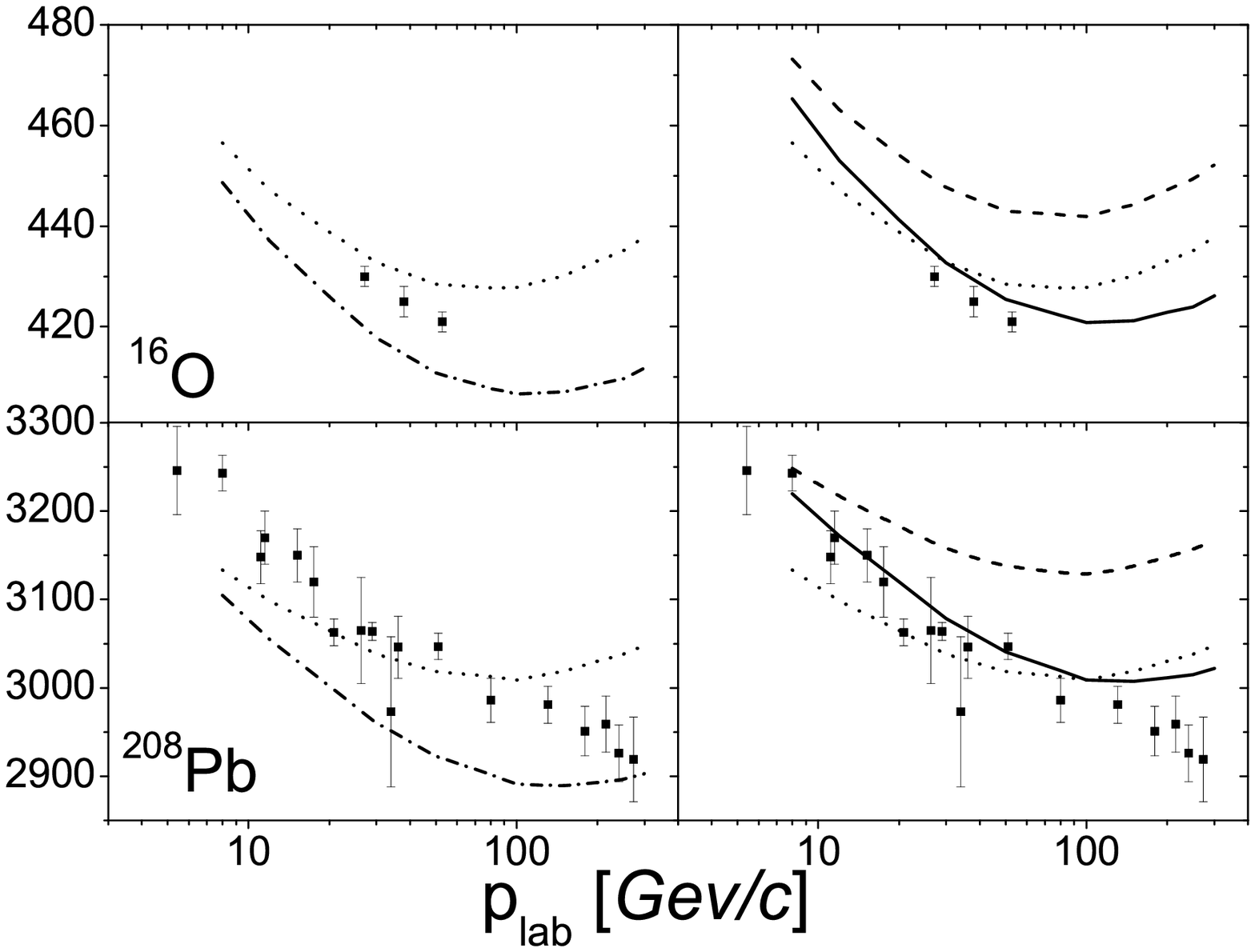}
}
\caption{The neutron-nucleus total cross section at high energies. LEFT PANELS:
  Glauber ($\sigma_G$;\,\,{\it dots}) and Glauber plus Gribov
  inelastic shadowing ($\sigma_G$ + $\Delta\sigma_{in}$;\,\, {\it dot-dashes}).
   RIGHT PANELS:
  Glauber ($\sigma_G$;\,\,{\it dots}), Glauber plus SRC ($\sigma_G$ + $\sigma_{SRC}$;\,\,
   {\it dashes}) and
  Glauber plus SRC  plus
  Gribov
  inelastic shadowing
  ($\sigma_G$ + $\sigma_{SRC}$ + $\Delta$ $\sigma_{in}$;\,\, {\it full}).
For references to the experimental data see
  \cite{murthy}. (After Ref. \cite{acmmp})}
  \label{Fig7}
\end{figure}
between a proton and a neutron. We found that in medium-weight nuclei
$P_{pp} \simeq 10\%$ and $P_{pn} \simeq 90 \%$ in agreement with the findings of Ref.
\cite{pia} obtained from the reactions $^{12}C(e,e'pn)$ and $^{12}C(e,e'pp)$.
The effects of various CM motion configurations on the value of $P_{pN}$ are being investigated
\cite{ACM1}.

\section{SRC and  High-Energy Scattering}

Nowadays the interpretation of high precision
particle-nucleus and nucleus-nucleus scattering experiments at
 high energies, aimed at investigating the state of
matter at short distances, should require in principle also a
consideration  of possible effects from NN correlations, which are usually disregarded
in Glauber-type approaches which consider  only the first term
({\it single density approximation})
of the exact  expansion
\beqy
\left|\Psi({\bf r}_1,...,{\bf r}_A)\right|^2=
\prod_{j=1}^A\rho({\bf r}_j)
      +\sum_{i<j=1}^A\Delta({\bf r}_i,{\bf r}_j)
      \prod_{k\neq(ij)}^A
            \rho_1({\bf r}_k)+ \dots
\label{psiquadro}
\eeqy
where
 $\Delta({\bf r}_i,{\bf r}_j)\,=\,
  \rho_2({\bf r}_i,{\bf r}_j)\,-\,
  \rho_1({\bf r}_i)\,\rho_1({\bf r}_j)$, yielding
   $\int d{\bf r}_2\Delta({\bf r}_1,{\bf r}_2)\,=\,0$.

In Ref. \cite{acmmp,acmmp1} the elastic, quasi-elastic and total nucleon-nucleus
cross sections  at high energies  have been calculated within Glauber approach taking
also into account all correlation terms in the expansion (\ref{psiquadro}) using
density matrices from Ref. \cite{acmprc2005}. The effects of correlations and Gribov inelastic
shadowing \cite{karma} on the total neutron-nucleus cross section,
which has been measured with high
precision, are given by
\be
\sigma_{tot}=2 Re \int d{\bf b}\,\left[1\,-\,\int \prod_{j=1}^A
d{\bf r}_j\,[1-\Gamma({\bf b}-{\bf s}_j)]\cdot
\left|\Psi_0({\bf r}_1,...,{\bf r}_A)\right|^2\,\right]
\label{crosstot}
\ee
 The results of calculations are  shown in Fig. \ref{Fig7},  which clearly
 exhibits the  role of SRC.

\section{Conclusions}
We have shown that a large wealth of different experimental data
concerning medium and high energy scattering off nuclei can be
interpreted within a framework which includes a proper treatment of
initial state short range correlations and final state
interactions. In the former, tensor and isospin-tensor correlations
appear to be the essential ingredients for a correct description
of one- and two-nucleon  momentum distributions both in few-body
systems and complex nuclei. As for the latter, Glauber multiple
scattering and the generalized eikonal approximation appear to be
a satisfactory frameworks for the description of the final state
interaction. At high values of the missing momentum both the
undistorted and distorted one-nucleon  and relative
momentum distributions strikingly resemble the same quantities
pertaining to deuteron. Finally, according to our results, the
effects of SRC on high energy scattering processes, if properly
treated,  should not be overlooked.
\bibliographystyle{aipproc}

\end{document}